\documentclass{aa}
\usepackage{graphicx}
\usepackage{natbib}
\usepackage{epstopdf}
\bibpunct{(}{)}{;}{a}{}{,}   
\newcommand{\lesssim}{\mathrel{\hbox{\rlap{\hbox{\lower4pt\hbox{$\sim$}}}\hbox{$<$}}}}
\newcommand{\grsim}{\mathrel{\hbox{\rlap{\hbox{\lower4pt\hbox{$\sim$}}}\hbox{$>$}}}}
\newcommand{\gammaeq}{\gamma_{\rm eq}}
\newcommand{\gammatrap}{\gamma_{\rm tr}}
\newcommand{\gammaexp}{\gamma_{\rm cat}}

\newcommand{\gammac}{\gamma_{\rm c}}

\newcommand{\taus}{\tau_{\rm s}}
\newcommand{\taup}{\tau_{\rm p}}

\newcommand{\eqb}{\begin{eqnarray}}
\newcommand{\eqe}{\end{eqnarray}}

\newcommand{\diff}{{\rm d}}
\newcommand{\sigmaT}{\sigma_{\rm T}}

\newcommand{\nelec}{n_{\rm e}}
\newcommand{\tauT}{\tau_{\rm T}}
\newcommand{\xabs}{x_{\rm a}}
\newcommand{\ghat}{\hat{\gamma}}

\newcommand{\doppler}{{\cal D}}
\newcommand{\Dten}{{\cal D}_{10}}
\newcommand{\nuabs}{\nu_{\rm abs}}
\newcommand{\nuGHz}{\nu_{\rm GHz}}
\newcommand{\nusynch}{\nu_{\rm s}}

\newcommand{\nufourteen}{\nu_{{\rm max,}14}}

\newcommand{\TB}{T_{\rm B}}
\newcommand{\PKSfifteennineteen}{PKS~1519~$-$273} 
\newcommand{\PKSofourofive}{PKS~0405~$-$385} 
\newcommand{\Jeighteennineteen}{J~1819~$+$3845}

\begin{document}
\title{The inverse Compton catastrophe and high brightness 
temperature radio sources}
\author{O.~Tsang
\and J.~G.~Kirk
}
\offprints{O.~Tsang, \email{olivia.tsang@mpi-hd.mpg.de}}

\institute{Max-Planck-Institut-f\"ur Kernphysik, Saupfercheckweg 1, 
D-69117 Heidelberg, Germany}

\date{Received \dots / Accepted \dots }
\abstract {
The inverse Compton catastrophe is the dramatic rise
in the luminosity of inverse-Compton scattered photons 
predicted to occur when the synchrotron brightness
temperature exceeds a threshold value, 
usually estimated to be $10^{12}\,$K.
However, this effect appears to be in contradiction with 
observation because: 
(i) the threshold is substantially 
exceeded by several intra-day variable 
radio sources, but the inverse Compton emission is not observed,
(ii) powerful, extra-galactic 
radio sources of known angular size do not appear to 
congregate close to the predicted maximum brightness temperature.}
{
We re-examine the parameter space available to 
synchrotron sources using a non-standard electron distribution, 
in order to see whether the revised 
threshold temperature is consistent with the data.}
{
We apply the 
theory of synchrotron radiation to a population of monoenergetic
electrons. The electron distribution and the population of each generation 
of scattered photons are computed using spatially averaged 
equations. The results are formulated in terms of the 
electron Lorentz factors that 
characterise sources at
the threshold temperature and sources in which the particle and 
magnetic field energy density are in equipartition.}
{
We confirm our previous finding that intrinsic brightness temperatures 
$T_{\rm B}\sim10^{14}\,$K can occur without 
catastrophic cooling. We show that substantially higher temperatures cannot 
be achieved either in transitory solutions or in solutions
that balance losses with 
a powerful acceleration mechanism. 
Depending on the observing frequency, we find 
strong cooling can set in at a range of threshold temperatures and 
the imposition of the additional 
constraint of equipartition between particle and 
magnetic field energy is not warranted by the data.}
{
Postulating a monoenergetic electron distribution, which approximates one 
that is truncated below 
a certain Lorentz factor ($\gamma_{\rm min}$), alleviates several
theoretical difficulties associated with the inverse Compton catastrophe,
including anomalously high brightness temperatures and the apparent
lack of clustering of powerful sources at $10^{12}\,$K.}
 
\titlerunning{The inverse Compton catastrophe}
\maketitle

\section{Introduction}
\label{intro}

When interpreting the synchrotron spectrum of powerful extra-galactic radio
sources, it is usual to assume that the underlying electron distribution has a
power-law form $\diff N/\diff\gamma\propto\gamma^{-q}$.  
For a homogeneous source, the spectrum peaks at the point
$\nu=\nuabs$, where the optical depth to synchrotron self-absorption 
is of the order of unity, and
above this frequency the intensity falls off as $I_\nu\propto
\nu^{-(q-1)/2}$.
Because more electrons become effective at
absorbing the radiation as the frequency decreases, the optically
thick part of the spectrum is not of the Rayleigh-Jeans type, but has 
instead
$I_\nu\propto\nu^{5/2}$, independent of the power-law index of the underlying
distribution, (provided $q>1/3$). 
Correspondingly, the brightness
temperature peaks at $\nu\approx\nu_{\rm abs}$, falling off as $\nu^{1/2}$ to
lower and as $\nu^{-(q+3)/2}$ to higher frequencies. In this case,
\citet{kellermannpaulinytoth69} found that the 
ratio of the luminosity $L_{\rm IC}$ 
carried away by inverse Compton
scattered photons to that of synchrotron photons $L_{\rm s}$ is
\eqb
{L_{\rm IC}\over L_{\rm s}}&=& \left({T_{\rm B}\over 
T_{\rm thresh}}\right)^{5}
\left[1+\left({T_{\rm B}\over T_{\rm thresh}}\right)^{5}\right]
\label{readheadcatastrophe}
\eqe
where $T_{\rm B}$ is the intrinsic 
brightness temperature at the peak of the radio
emission and $T_{\rm thresh}\approx10^{12}\,$K, depending somewhat on the
parameter $q$ and the maximum frequency at which synchrotron radiation is
emitted, corresponding to an assumed cut-off in the power-law electron
spectrum \citep{readhead94}. 

The rapid rise in total luminosity implied by Eq.~(\ref{readheadcatastrophe})
when $T_{\rm B}$ exceeds $T_{\rm thresh}$ is called the \lq\lq inverse Compton
catastrophe\rq\rq. The implied energy requirement of a $1\,$Jy source at $z=1$
is $\sim10^{43}{\rm ergs}\,{\rm s}^{-1}$ at $T_{\rm B}=10^{12}\,$K, and a
prohibitive $\sim10^{53}{\rm ergs}\,{\rm s}^{-1}$ at $T_{\rm B}=10^{13}\,$K at
1GHz. A source of $1\,$GHz photons with magnetic field $\approx1\,\mu$G,
boosts a radio photon into the X-ray band in a single inverse Compton
scattering. At $T_{\rm B}=10^{12}\,$K this would imply a nanoJansky X-ray
flux, not untypical of strong extragalactic sources. However, at $T_{\rm
B}=10^{13}\,$K Eq.~(\ref{readheadcatastrophe}) predicts an X-ray flux at the
milliJansky level, in contradiction with observation
\citep[]{fossatietal98,sambrunaetal00,tavecchioetal02,padovanietal04,guainazzietal06}.

Despite this, several sources that display intra-day variability (IDV) in
their radio emission have an implied brightness temperature that exceeds
$10^{12}\,$K by several orders of magnitude, if the observed variability is
intrinsic \citep[e.g.,][]{krausetal03}. Even if the variability is caused by
scintillation, the implied brightness temperature can still greatly exceed
$10^{12}\,$K for some sources \citep{wagnerwitzel95}. Currently, the most
extreme example is the source PKS~0405-385. This source displays diffractive
scintillation \citep{macquartdebruyn05}, which places an upper limit on its
angular size that corresponds to a brightness temperature of
$2\times10^{14}\,$K.  These sources are generally assumed to be {\em beamed},
i.e., to be in relativistic motion towards the observer
\cite[e.g.,][]{rees66,jonesburbidge73,singalgopalkrishna85}. In this case the
intrinsic temperature is lower than that deduced for a stationary source by a
factor of ${\cal D}^3$ (for intrinsic variability) or ${\cal D}$ (for
scintillating sources). (Here the Doppler factor ${\cal
D}=\sqrt{1-\beta^2}/(1-\beta\cos\theta)$ with $\beta c$ the source velocity
and $\theta$ the angle between this velocity and the line of sight.)
Nevertheless, the observed brightness temperatures are too high to be
accounted for by Doppler factors similar to those estimated from observations
of superluminal motion \citep{cohenetal03}.

A second problem arises with powerful sources whose angular extent can be
measured directly.  In an analysis of high brightness temperature radio
sources in which Doppler beaming is thought to be absent, \citet{readhead94}
measured a brightness distribution that cuts off at $10^{11}\,$K; one order of
magnitude lower than the inverse Compton limit.  This appears consistent with
observations of a sample of 48 sources showing superluminal motion
\citep{cohenetal03}, in which it was found that the intrinsic brightness
temperature cluster around $2\times10^{10}$ K. \citet{readhead94} argued that
an apparent maximum brightness temperature significantly lower than
$10^{12}\,$K could not be caused by catastrophic Compton cooling. Instead, he
suggested that sources are driven towards equipartition between their magnetic
and particle energy contents. Assuming, in addition, that observations are
taken at the peak of the synchrotron spectrum, and that the electron
distribution is a power-law, he showed that the expected distribution of
brightness temperatures was consistent with that observed.

In this paper, we re-examine these two problems
assuming that the source contains a monoenergetic electron distribution 
instead of the conventional power-law. Although this assumption 
appears at first sight highly 
restrictive, the form of the synchrotron emissivity means that 
under some circumstances
such a distribution provides a good approximation to several more commonly 
encountered cases, including that of a conventional 
power-law distribution that is truncated to lower energy at a 
Lorentz factor $\gamma_{\rm min}$. Such distributions have been proposed 
in connection with radio sources for 
a variety of reasons: 
the absence of low energy electrons can account for the lack
of Faraday depolarisation in parsec-scale emission regions
\citep{wardle77,jonesodell77} and has recently been discussed in
connection with statistical trends in the observed distribution of
superluminal velocities as a function of observing frequency and
redshift \citep{gopal-krishnaetal04}. 
Also, \cite{blundelletal06} recently
examined the radio and x-ray emission from the lobe regions of a giant
radio galaxies 6C 0905$+$3955, and deduced a low energy cutoff of
the relativistic particles in the hotspots of $\gamma_{\rm
min}\sim10^4$.

In Sect.~\ref{synchrotron} we use standard theory to discuss 
the general properties of the synchrotron spectra emitted by a 
homogeneous source. A set of spatially averaged equations describing the 
evolution of the electron Lorentz factor and both the synchrotron and 
the associated inverse Compton scattered emission 
is presented in Sect.~\ref{spatiallyaveraged}. 
Having identified in these equations the threshold for 
the inverse Compton catastrophe, we discuss the parameter space 
available to stationary solutions in Sect.~\ref{stationary}. Here we 
confirm the results reported in \cite{kirktsang06}, indicating that
temperatures considerably in excess of $10^{12}\,$K are permitted. 
We also show that in the case of resolved sources, 
the onset of catastrophic cooling occurs over a wide range
of temperatures, consistent with the observed temperature range.
Finally, we address in Sect.~\ref{timedependent} the 
suggestions by \cite{slysh92} that extremely high brightness 
temperatures can be achieved in nonstationary sources either by injecting 
electrons at high energy, or by balancing their cooling 
against a powerful acceleration mechanism. A summary of our conclusions 
is presented in Sect.~\ref{conclusions}.

\section{Synchrotron spectra}
\label{synchrotron}
We consider a homogeneous source region 
characterised by a single spatial scale $R$, that contains
monoenergetic electrons and possibly positrons of Lorentz factor
$\gamma$ and number density $\nelec$ immersed in a magnetic field
$B$. Expressions for the synchrotron emissivity and absorption
coefficients can be found in many excellent texts 
(e.g., \citet[Chapter 6]{rybickilightman79},
and \citet[chapter 18]{longair92}) and are
summarised in our notation in Appendix~\ref{synchformulae}.

\begin{figure}
\resizebox{\hsize}{!}{%
\includegraphics{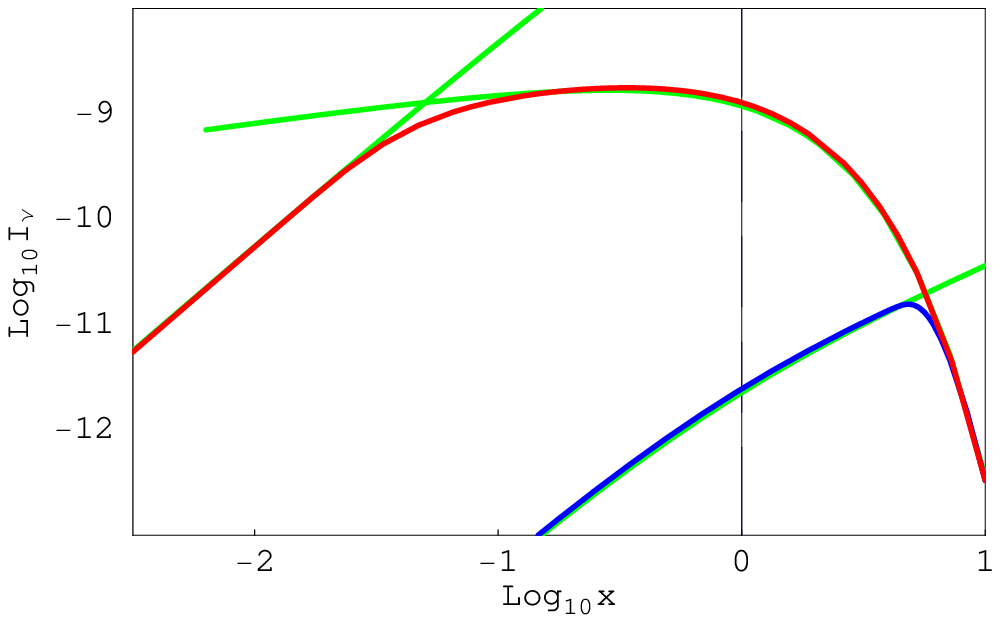}}
\resizebox{\hsize}{!}{%
\includegraphics{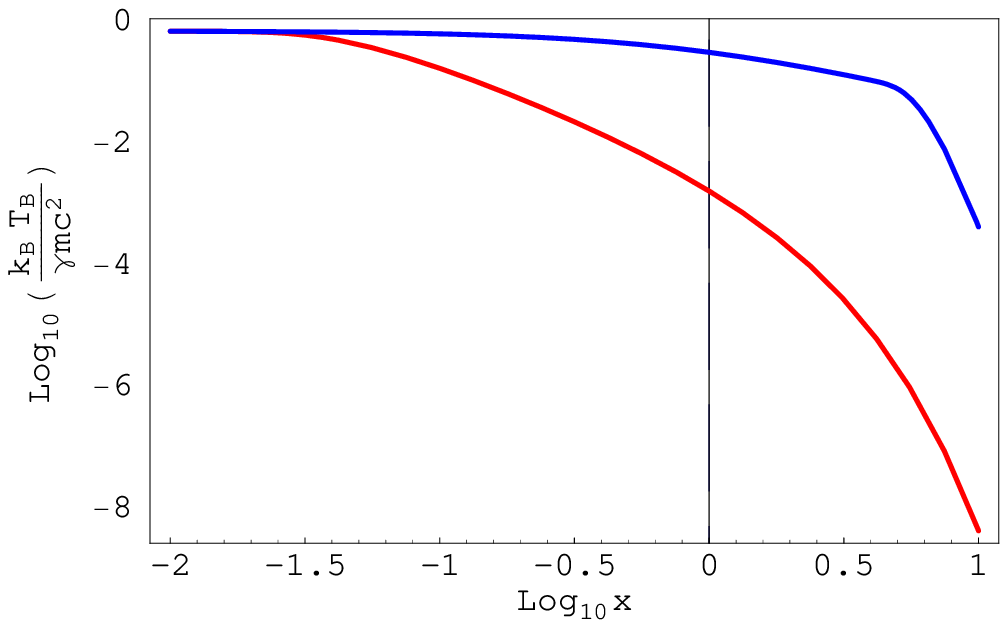}}
\caption{%
\label{synchspectra}
The synchrotron spectra (upper panel) and brightness temperatures (lower
panel) of sources with monoenergetic electrons in the case of strong (blue)
and weak (red) absorption. The green curves show the optically thick
($I_\nu=S_{\nu}$, see Appendix~\ref{synchformulae}) and optically thin
($I_\nu=\taus S_{\nu}$) approximation.  In the upper panel, $I_\nu$ is in
arbitrary units, and in the lower, the brightness temperature is normalised to
the energy of the electron.  The abscissa $x$ is the ratio of the frequency to
the characteristic synchrotron frequency of the electrons $\nu_{\rm s}$.  The
blue (red) curves correspond to a source which has an optical depth of unity
to synchrotron self-absorption at $x\approx5$ ($x\approx0.05$).  For ease of
display, the upper panel compares sources with equal flux at high frequency,
whereas the lower compares sources with equal flux at low frequency.}
\end{figure}

For any given source there exists a frequency $\nuabs$ below which
absorption is important. Since $B$ and $\gamma$ also define a
characteristic synchrotron frequency $\nusynch$ (see
Eq.~\ref{nusynchdef}), the sources we consider can be divided into two
categories: those with {\em weak absorption} in which
$\nuabs<\nusynch$ and those with {\em strong absorption}
$\nuabs>\nusynch$. Note that this division is independent of the 
observing frequency, since it relates only to intrinsic source 
properties.  The synchrotron spectra that emerge in these two cases
are quite
different, and are illustrated in Fig.~\ref{synchspectra}. A feature
they have in common is that the low energy spectrum has the
Rayleigh-Jeans form $I_{\nu}\propto \nu^{2}$, where $I_{\nu}$ is the
specific intensity at frequency
$\nu$. This property contrasts with the $\nu^{5/2}$
dependence of $I_{\nu}$ at low frequencies of a source containing a
power-law distribution of electrons. The reason is that a power-law
distribution contains cold (low energy) electrons that contribute to
the absorption at low frequencies.

The brightness temperature, $T_{\rm B}=c^2 I_{\nu}/(2\nu^2 k_{\rm B})$, where
$k_{\rm B}$ is Boltzmann's constant, is a function of frequency and is also
illustrated in Fig.~\ref{synchspectra}. At low frequency, it attains its
maximum value roughly in \lq\lq equilibrium\rq\rq\ with the electrons: $T_{\rm
B, max}= 3\gamma m c^2/4k_{\rm B}$, then decreases monotonically to higher
frequencies. In the case of weak absorption, $T_{\rm B, max}\propto\nu^{-5/3}$
for $\nuabs<\nu<\nusynch$, and then cuts off exponentially as
$\nu^{-3/2}\exp\left(-\nu/\nusynch\right)$ once $\nusynch$ is exceeded. In
strongly absorbed sources, the brightness temperature remains almost constant
until the frequency exceeds $\nusynch$ upon which it falls off as $\nu^{-1}$
until the source becomes optically thin, after which the exponential cut-off
$T\propto \nu^{-3/2}\exp\left(-\nu/\nusynch\right)$ takes over.

Although four parameters ($\gamma$, $\nelec$, $B$ and $R$)
are needed to define a source model, the division between strong and weak 
absorption is simple. It occurs at a critical Lorentz factor $\gammac$
given by (see Eq.~\ref{gammacdef})
\eqb
\gammac&=&324\times \left({\nelec\over 1\,\textrm{cm}^{-3}}\right)^{1/5}
\left({R\over 1\,\textrm{kpc}}\right)^{1/5}
\left({B\over 1\,\textrm{mG}}\right)^{-1/5}
\label{strongweak}
\\
\noalign{\hbox{or, equivalently,}}
\gammac&=&4451\times \tauT^{1/5}
\left({B\over 1\,\textrm{mG}}\right)^{-1/5}
\label{strongweak2}
\eqe
where $\tauT=\nelec R\sigmaT$ is the Thomson optical depth of the source. 
Strong absorption occurs for low Lorentz factors $\ghat=\gamma/\gammac<1$
and weak absorption for high Lorentz factors $\ghat>1$. 
If the Lorentz factor $\gamma$ is held constant, the strong absorption 
regime may be reached from the weak by increasing 
$\tauT$ at constant $B$, or by {\em decreasing} $B$ at constant $\tauT$.   

In his model of high-brightness temperature sources, 
\citet{slysh92} considered the strong absorption
case. The most important property of the assumed distribution in this
case is the lack of high energy electrons: the addition of a
population of cold electrons, which would correspond to a power-law
distribution truncated to higher Lorentz factors, would reduce the
brightness temperature of the source at $x<1$ (in
Fig.~\ref{synchspectra}) but would not significantly influence this quantity, 
for $x>1$.

On the other hand,
\citet{crusius-waetzel91} and \citet{protheroe03} considered weak
absorption, where the key property of the model distribution is the
absence of low energy electrons. In this case, the 
monoenergetic model is a good approximation to 
a power-law distribution truncated to lower electron
energies at $\gamma=\gamma_{\rm min}$. The addition of a 
high-energy power-law tail affects the spectrum at $x>1$, but 
does not change the maximum brightness temperature achieved at
$x\lesssim1$. Furthermore, 
the truncation need not be sharp: provided the
opacity at low frequencies is dominated by the contribution of 
electrons with $\gamma\approx\gamma_{\rm min}$, the monoenergetic 
approximation is good. This is the case if, for 
$\gamma<\gamma_{\rm min}$, the spectrum is sufficiently hard: 
$\diff N/\diff\gamma\propto \gamma^{-q}$ with $q\le1/3$. In particular, 
the low energy tail of a relativistic Maxwellian 
distribution ($q=-2$) falls into this category. 

In contrast to the pure power-law distribution, where the
self-absorption turnover is strongly peaked, the 
emission of a weakly absorbed source --- shown in red in 
the upper panel of 
Fig.~\ref{synchspectra} --- is flat over nearly two decades in frequency.
It therefore provides a natural
explanation of compact flat-spectrum sources, eliminating the need to
appeal to a \lq\lq cosmic conspiracy\rq\rq\ 
behind the superposition of peaked spectra
from different parts of an inhomogeneous source \citep{marscher80}. 

For the treatment of inverse Compton scattering, it is
necessary to evaluate the the energy density $U_{\rm s}$
in synchrotron photons in a given source. To do this,  
$I_\nu$ must be integrated over angles and 
over frequency. The result depends on the geometry and 
optical depth as well as the position
within the source. An average value can be estimated by 
introducing a geometry dependent factor, $\zeta$, defined according to:
\eqb
U_{\rm s}&\approx&{4\pi \zeta\over c}\int_0^\infty \diff\nu 
\left<I_\nu\right>
\label{xidef}
\eqe where $\left<I_\nu\right>$ is conveniently taken to be the specific
intensity along a ray path that is within the source for a distance $R$ and is
perpendicular to the local magnetic field. \cite{protheroe02} has evaluated
$\zeta$ for several interesting special cases. For a roughly spherical source,
it is of the order of unity. Below, we show that the choice $\zeta=2/3$ is
consistent with our spatially averaged treatment of the kinetic equations. The
dominant contribution to the integral over the spectrum arises from photons of
frequency close to $\nusynch$ in the case of weak absorption, and close to
$\nuabs$ in the case of strong absorption. Using this approximation, for weak
absorption ($\ghat>1$): 
\eqb U_{\rm
s}&\approx&4.1\times10^{-6}\gamma^2\zeta\left({B^2\over8\pi}\right)
\left({\nelec\over 1\,\textrm{cm}^{-3}}\right) \left({R\over
1\,\textrm{kpc}}\right) \\ \noalign{\hbox{or, equivalently,}} U_{\rm
s}&\approx&2\gamma^2\tauT\zeta\left({B^2\over8\pi}\right)
\label{uweakabs}
\eqe
and for strong absorption ($\ghat<1$):
\eqb
U_{\rm s}&\approx&8.9\times10^{-18}\gamma_c^7\zeta\left({B^2\over8\pi}\right)
\left({B\over 1\,\textrm{mG}}\right)\left(\ln\ghat\right)^2
\label{ustrongabs}
\eqe
An approximation that is accurate for all values of the optical depth
is given in Eq.~(\ref{simpleapprox}).
\section{Spatially averaged equations}
\label{spatiallyaveraged}

An approximate, spatially averaged 
set of equations governing the energy balance of particles and
synchrotron radiation in a source can be found following the 
approach of \citet{lightmanzdziarski87} 
and \citet{mastichiadiskirk95}. In
terms of the time-dependent synchrotron radiation energy density $U_0(t)$ 
one can write:
\eqb
{\diff U_0\over\diff t} + c \left<\alpha_\nu\right> U_0 + {c\over R} U_0&=&
\left<j_\nu\right>
\label{synceqa}
\eqe
The second and third terms on the left-hand side of this equation represent the
rate of energy loss by 
the radiation field due to synchrotron self-absorption and
escape through the source boundaries; the right-hand side is the power put into
radiation by the particles. The angle brackets indicate a
frequency and angle average, but, within this spatially-averaged treatment, 
an exact calculation of the frequency average is unnecessary; it suffices to
replace the absorption coefficient by its value where the 
energy density of the synchrotron spectrum peaks i.e., at $\nu=\nusynch$ 
in the case of weak absorption and $\nu=\nuabs$ 
in the case of strong
absorption. In terms of the optical depth to synchrotron absorption 
at this point, $\taup\leq1$, the equation becomes:
\eqb
{\diff U_0\over\diff t} + {c\over R}\left(1+\taup\right)U_0 &=&
\left<j_\nu\right>
\label{synceqb}
\eqe
The right-hand side of this expression 
can now be found by demanding it gives the
correct steady solution at both large and small optical depth. The resulting
equation is:
\eqb
{\diff U_0\over\diff t} + 
{c\over R}\left(1+\taup\right)\left[U_0(t)-U_{\rm s}(\gamma)\right] &=&0
\label{synceq}
\eqe
where $U_{\rm s}$ is the steady-state synchrotron radiation energy
density, evaluated according to Eq.~(\ref{xidef}), with an appropriate value of
the parameter $\zeta$. 

The corresponding equation for the particles that takes into account
synchrotron absorption and emission as well as an acceleration term 
takes the form
\eqb
n_{\rm e}mc^2
{\diff \gamma\over\diff t} &=& {c\over R} \taup U_0 
- {c\over R}\left(1+\taup\right)U_{\rm s}
+ 
a\,eBc n_{\rm e}
\label{electroneq}
\eqe
The first term on the right-hand side of Eq.~(\ref{electroneq}) 
is the power taken from the radiation field by self-absorption
and the second term is that returned to it ---  both of these appear in
Eq.~(\ref{synceq}).  The third term
describes the energy input by particle acceleration. The
particular scaling used follows that of \citet{slysh92}, 
and models a generic first-order
Fermi process. For $a$ independent of $\gamma$, the 
acceleration rate is proportional to the gyro
frequency, and for $a=1$
it equals this value. The acceleration timescale equals the 
crossing time of the source when $a=\gamma mc^2/(eBR)$.

Multiple inverse Compton scatterings 
can be accounted for as follows: First we label the 
photons present in the source according to how many scattering events they 
have
suffered after production by the synchrotron process. 
The energy density of these photons is denoted by $U_i$
Thus, $i=0$ corresponds to photons emitted by the synchrotron process which
have not undergone a scattering, and the corresponding energy density 
is governed by Eq.~(\ref{synceq}). 
Assuming the source is optically thin to Thomson
(or Compton) scattering, the dominant loss mechanism for 
the energy density of photons
belonging to a given generation $i\ge1$ is escape from the source, rather than
conversion to the $i+1$'th generation. In this case, we can write 
for the time-dependence of $U_i$:
\eqb
{\diff U_i\over\diff t}+ {c\over R}U_i &=& Q_i
\label{timedepui}
\eqe
where $Q_i$ is the rate per
unit volume at which energy is transferred into photons of 
the $i$'th generation by inverse Compton scattering, for $i\ge1$, or by
synchrotron radiation for $i=0$. 

If the inverse scattering process proceeds in the Thomson regime
a simple expression can be found for $Q_i$. 
However, as $i$ increases, $h\nu_i$ 
also increases, eventually becoming comparable to the electron
energy when viewed in its rest frame. When this
happens, 
Klein-Nishina modifications to the Thomson 
cross section become important, reducing the value $Q_i$. 
We take approximate account of this effect by limiting the number of
scatterings to $N_{\rm max}$, and using the Thomson approximation 
to evaluate $Q_i$ for $i\le N_{\rm max}$. 
In this case, the average energy 
of a scattered photon of the $i'$th generation is $\nu_i=4\gamma^2
\nu_{i-1}/3$ and the rate of such scatterings in unit volume of the source is
$\nelec\sigmaT c U_{i-1}/(h\nu_{i-1})$. Therefore 
\eqb
Q_i&=&\left\lbrace
\begin{array}{ll} 
\xi c
U_{i-1}/R 
&\textrm{for\ }1\le i\le N_{\rm max}\\
&\\
0&\textrm{for\ }i> N_{\rm max}
\end{array}\right.
\label{defqi}
\eqe
where the parameter $\xi$ is defined as
\eqb
\xi&=&{4\over3}n_{\rm e}\sigmaT R \gamma^2 
\nonumber\\
&=&
{4\gamma^2\tauT\over3}
\label{xipar}
\eqe

The appropriate value of $N_{\rm max}$ is chosen by requiring the 
average energy of the $N_{\rm max}$ generation 
of photons viewed in the electron rest frame
$\gamma(4\gamma^2/3)^{N_{\rm max}}h\nu_0$ to be less than 
the electron energy:
\eqb
N_{\rm max}&=&\textrm{floor}\left[{\ln\left(mc^2/h\nu_0\right)\over 
2\ln\gamma}+{1\over2}\right]  
\label{defnmax}
\eqe
For synchrotron radiation, Eq.~(\ref{synceq}) implies
\eqb
Q_0&=&{c\taup\over R}\left(U_{\rm s}-U_0\right) +{c\over R}U_{\rm s}
\label{defqzero}
\eqe
In the stationary case, $U_0=U_{\rm s}$, 
Eqs.~(\ref{defqi}) and (\ref{defqzero}) give $Q_1/Q_0=\xi$.
However, assuming scattering in the Thomson regime, the ratio of the 
energy lost by synchrotron radiation to that by inverse Compton scattering 
in the steady state equals 
the ratio of the energy density of the magnetic field to that of the
target photons $Q_0/Q_i=B^2/(8\pi U_{i-1})$, which, for $i=0$, implies
$U_{\rm s}=\xi\left(B^2/8\pi\right)$. Comparison with
Eq.~(\ref{uweakabs}) then confirms that the spatially averaged kinetic
equations are consistent with the choice $\zeta=2/3$ for the geometry dependent
factor. Finally, the electron equation~(\ref{electroneq}) acquires the 
additional loss
terms from inverse Compton scattering:
\eqb
\nelec mc^2{\diff \gamma\over\diff t} &=& -\sum_{i=0}^{N_{\rm max}} Q_i
+ 
a\,eBc \nelec
\label{electroneqb}
\eqe
 
The set of equations 
(\ref{timedepui}) and (\ref{electroneqb}) can be rewritten
by introducing the
total energy density of scattered radiation: 
\eqb
U_{\rm T}&=&\sum_{i=1}^{N_{\rm max}} U_i
\label{useries}
\eqe
Then, using
dimensionless variables according to $\hat{U}=
U\left(8\pi/B^2\right)$, 
$\hat{t}=tc/R$ and $\hat{Q}_i=8\pi c Q_i/(R B^2)$ 
one finds
\eqb
{\diff \hat{U}_{\rm T}\over \diff \hat{t}}+\left[1 - 
\xi\right]\hat{U}_{\rm T}&=&
\xi\left(\hat{U}_0 -\hat{U}_{N_{\rm max}}\right)
\label{expl}
\eqe
If $U_{N_{\rm max}}$ remains always negligibly small, then
all significant scatterings occur in the Thomson regime,
and the set of equations (\ref{timedepui}) (for $i=0$), (\ref{electroneqb}),
and (\ref{expl}) can be conveniently formulated in terms of
three characteristic values of the Lorentz factor: 
\eqb
{\diff \hat{U}_{\rm T}\over \diff \hat{t}}&=&-\left[1 - 
\left(\gamma/\gammaexp\right)^2\right]\hat{U}_{\rm T} +
\left(\gamma/\gammaexp\right)^2 \hat{U}_0 
\label{explosive}
\\
{\diff \hat{U}_0\over \diff \hat{t}}&=&-\hat{U}_0+\hat{Q}_0
\label{explosivea}
\\
{\diff\gamma\over\diff \hat{t}}&=&-\gammaeq\left[\hat{Q}_0+
\left(\gamma/\gammaexp\right)^2\hat{U}_{\rm T}\right] 
+\gammatrap a
\label{electroneqa}
\eqe 
where $\gammaeq$ is chosen so that there is equipartition between particle and
magnetic energy densities for $\gamma=\gammaeq$:
\eqb
\gammaeq&=&B^2/(8\pi n_{\rm e}m c^2) 
\label{defgeq}
\eqe
$\gammaexp$ is given by setting $\xi=1$
\eqb
\gammaexp&=&\sqrt{3\over4\tau_{\rm T}}
\label{defgcat}
\eqe
and $\gammatrap$ corresponds to the maximum Lorentz factor of a particle
that can be confined in the source, i.e., whose gyro-radius is less than $R$:
\eqb 
\gammatrap&=&eBR/(mc^2)
\eqe

The 
significance of $\gammaexp$ can be seen from the steady state solution of
Eqs.~(\ref{explosive}) and (\ref{explosivea}): $U_{\rm T}=U_{\rm
s}/\left(\gammaexp^2/\gamma^2-1\right)$.  For values of $\gamma$ that approach
$\gammaexp$ from below, the energy density in the radiation field, and, hence,
the luminosity diverge. Thus, under the assumption that all scatterings take
place in the Thomson limit, no stationary solutions can be found for 
\eqb
\gamma\ge\gammaexp
\label{catastrophe}
\eqe
This phenomenon is the nonrelativistic or \lq\lq Thomson\rq\rq\ manifestation
of the {\em Compton catastrophe} described in the Introduction.  
In the weak absorption limit, $\hat{U}_{\rm s}=\gamma^2/\gammaexp^2$, 
confirming the well-known result that
the Compton catastrophe sets in when the energy density in synchrotron 
photons exceeds
the magnetic energy density. However, this result does not
apply to the case of strong absorption, where we find
$\hat{U}_{\rm s}\sim\gammaexp^5/\gammac^5\ll1$. In this regime, 
the synchrotron radiation energy density can be much smaller than the energy
density in the magnetic field at the point where catastrophic cooling sets
in. Physically, the scattered photons feed on each other to produce the
catastrophe in this regime, and do not require a substantial synchrotron photon
density. 
In a realistic model, the divergence of the luminosity is prevented by
Klein-Nishina effects, that effectively truncate the series in
Eq.~(\ref{useries}). For example, if $T_{\rm B, max}=10^{12}\,$K, at an
observing frequency of $1\,$GHz, so that $\gamma\approx 200$, then, from Eq.~(\ref{defnmax}), the number
of terms contributing to the sum is $N_{\rm max}=2$.

\section{Stationary solutions}
\label{stationary}
\subsection{Intra-day variable sources}
For comparison with observations of intra-day variable sources, it is
convenient to formulate the expression for the 
specific intensity given in Eq.~(\ref{specintappendix}) in terms of 
quantities accessible to observation. Expressing the result in terms of the 
observed (at $z=0$) brightness temperature, we find, in the case of weak
absorption, and at low frequency ($\nu\ll\nu_{\rm s}$)
\eqb
T_{\rm B}\!=\!
1.2\!\times\!10^{14}\!
\left(\!{\doppler_{10}^{6}\xi\over(1\!+\!z)^6}\!\right)^{\!1/5}\!\!
\left(\!{1\!-\textrm{e}^{-\taus}\over\taus^{1/5}}\!\right)\!
\nu_{{\rm max}14}^{2/15}
\nuGHz^{-1/3}\,\,
\textrm{K}
\label{bright1}
\eqe
\citep{kirktsang06}
where
$\doppler=10\Dten$ is
the Doppler boosting factor, $z$ is the redshift of the host galaxy, $\taus$
is the optical depth of the source at the observing frequency $\nu=\nuGHz$
GHz, and the characteristic synchrotron frequency of the electrons is
$\nusynch=\nufourteen\times10^{14}$ Hz.

According to Eq.~(\ref{bright1}), brightness temperatures 
of $T_{\rm B}\approx10^{13}\,$K, such as observed in the sources 
\PKSfifteennineteen\ 
and \PKSofourofive\ 
\citep{macquartetal00,rickettetal02}
can be understood within a simplified homogeneous 
synchrotron model in which $\xi\lesssim1$, implying a relatively modest inverse
Compton luminosity, i.e., no catastrophe. Even the extremely compact source
\Jeighteennineteen, which has $T_{\rm B}\grsim2\times10^{14}\,$K can be
accommodated in a catastrophe-free model 
provided the Doppler factor is greater than about 15. 
In each case, a hard spectrum is predicted, extending to
$\nufourteen\times10^{14}$ Hz. 
Although the dependence of the brightness
temperature on this parameter is quite weak, 
simultaneous observations in the radio to IR and optical 
\citep{ostoreroetal06}
have the potential
to rule out this explanation on a source by source basis.
 
\subsection{Resolved sources}
In a seminal paper, \cite{readhead94} discussed the distribution in brightness
temperature of a sample of powerful sources whose angular size could either 
be measured directly, or 
constrained by interplanetary scintillation. In discussing these objects
several simplifications must be made, even 
within the context of a homogeneous synchrotron model.

Firstly, in the two low frequency samples ($81.5\,$MHz and $430\,$MHz)
considered by \citet{readhead94}, the emission is thought to be almost
isotropic.  Doppler boosting is then unimportant and can be
neglected. Secondly, these sources are not very compact; their extension on
the sky is typically between $0.1$ and $1\,$arcsec. Therefore, for our
discussion we fix the linear extent $R$ of the source to $1\,$kpc,
corresponding to an angular size of approximately $0.2\,$arcsec at redshift
$z=1$.  This leaves three parameters needed to specify the source model: the
magnetic field strength $B$, the electron density $n_{\rm e}$ and the Lorentz
factor $\gamma$ of the electrons.  In order to clarify the physics of a
source, we transform from the parameter set $(B,n_{\rm e},\gamma)$ to the
characteristic Lorentz factors $\gammaeq$ and $\gammaexp$ defined in
Eqs.~(\ref{defgeq}) and (\ref{defgcat}).  Our basic parameter set is therefore
$(\gammaeq,\gammaexp,\gamma$).
Finally, in order to 
display on a two-dimensional figure source properties such as 
brightness temperature and spectral slope at a particular frequency, 
we consider a slice through this three dimensional parameter space,
selecting parameters such that the particle and
magnetic energy densities are in equipartition: $\gamma=\gammaeq$.

\begin{figure}
\resizebox{\hsize}{!}{%
\includegraphics[width=8 cm]{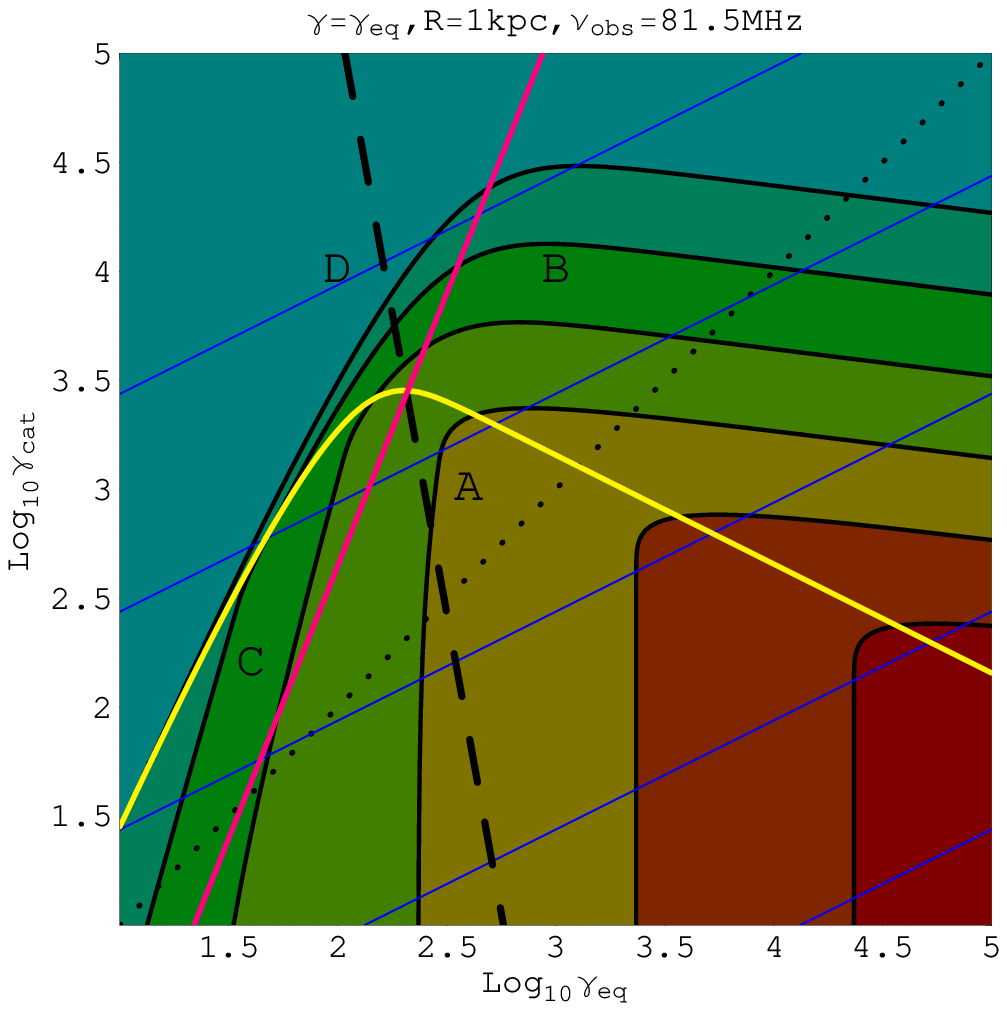}}
\resizebox{\hsize}{!}{%
\includegraphics[bb=90 30 370 95, clip]{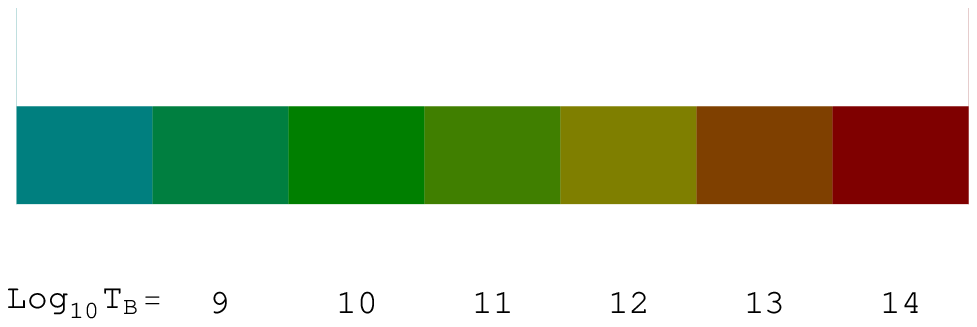}}
\caption{\label{figureeq} 
The brightness temperature as a function of $\gammaeq$ and $\gammaexp$
assuming equipartition between the magnetic and particle energy densities and
a source size $1\,$kpc.  Black contour lines indicate
$\log_{10}(T/\textrm{Kelvin})= 9$, $10$, $11$, $12$, $13$ and $14$.
The red line is the locus of points at which the characteristic synchrotron frequency
of the emitting particles is $81.5\,$MHz, the yellow line shows where the
source has an optical depth of unity at this frequency. The dashed line
divides regions of strong absorption (to the left) from those of weak
absorption (to the right). The diagonal $\gammaeq=\gammaexp$ is shown as a
dotted line.  Contour lines of the magnetic field strength are shown in blue,
ranging from $\log_{10}(B/\textrm{Gauss})=-4$ to $0$ (in the bottom right-hand
corner).}
\end{figure}

The properties of source models on this slice are shown in the 
$\gammaeq$--$\gammaexp$ plane 
in Fig.~\ref{figureeq}.
This plane can immediately be divided into regions
of strong and weak absorption, as defined in
Eq.~(\ref{strongweak}). The boundary, drawn as a thick dashed line,
represents the locus of the points at which 
$\gammac=\gammaeq$.
Weakly absorbed sources lie towards higher 
$\gammaeq$ and $\gammaexp$ (i.e., the upper-right side) and strongly absorbed
sources towards lower $\gammaeq$ and $\gammaexp$ 
(i.e., the lower-left side). 
We  also show (in blue) contours of the 
magnetic field strength.  

The remaining source properties depend upon the choice of observing
frequency. In Fig.~\ref{figureeq} we take this to be
$81.5\,$MHz, corresponding to the low frequency sample discussed by
\cite{readhead94}.
In order to determine the spectral
slope of a given source, we plot as a yellow line the locus of points where
the observing frequency coincides with the frequency at which the optical
depth to absorption is unity, $\nu_{\rm abs}$. Sources that lie above this line
(on the side of larger $\gammaexp$) are optically thin at the chosen observing
frequency. In addition, the red line in Fig.~\ref{figureeq} 
gives the locus of points where the observing frequency equals
the characteristic frequency of synchrotron radiation $\nu_{\rm c}$. By
definition, the intersection point of these lines lies also on the boundary
between weak and strong absorption (the thick dashed line). 
The observing frequency lies
below $\nu_{\rm c}$ on the lower-right side of the red line. 
The colour shading gives the intrinsic brightness temperature at the chosen
observing frequency.

The two lines (yellow and red) divide the $\gammaeq$-$\gammaexp$-plane 
in Fig.~\ref{figureeq} into
four regions with differing spectral properties: in region~A, sources have a
Rayleigh-Jeans spectrum $I_\nu\propto\nu^2$, in region~B, the spectrum is that
of low frequency, optically thin synchrotron radiation $I_\nu\propto
\nu^{1/3}$, in region~C, it is close to $I_\nu\propto\nu$
\citep[see][]{slysh92} and in region~D it falls off exponentially
$I_\nu\propto \nu^{-1/2}\exp(-\nu/\nu_{\rm c})$. Consequently, flat spectrum
sources reside in region~B, preferentially close to the yellow line and in
region~C, preferentially close to the red line.

Sources that are in equipartition and lie below the threshold of the Compton
catastrophe are to be found in the upper left half of Fig.~\ref{figureeq},
above the dotted line on which $\gammaeq=\gammaexp$. The maximum brightness
temperature accessible to these sources occurs close to
$\gammaeq=\gammaexp=10^3$, and is approximately $10^{12.6}\,$K, in rough
agreement with the results of \citet{kellermannpaulinytoth69}, who, however,
did not assume their sources to be in equipartition. 
The brightest sources are weakly absorbed, (they 
lie to the right of the thick dashed line)
and have a magnetic field strength of a few milliGauss. Their optical depth to
synchrotron self-absorption lies close to unity at the observation frequency
(they lie close to the yellow line).

\citet{singalgopalkrishna85} first discussed the effects of 
the additional assumption of equipartition on bright sources and 
used it to estimate Doppler factors for rapidly variable sources. 
Later, \citet{readhead94} introduced the concept of an 
\lq\lq equipartition brightness temperature\rq\rq\ 
to explain the observation
that the temperature distribution of resolved sources 
appears to peak significantly
below $10^{12}\,$K. However, the crucial additional assumptions in his
treatment is that the source flux is measured at the \lq\lq synchrotron
peak\rq\rq, and that the electron distribution is a power-law in energy.
This 
implies that the opacity at a given frequency (e.g., at the synchrotron peak)
is dominated by those electrons with a corresponding 
characteristic frequency. 
In our model, in which the electron distribution is approximated
as monoenergetic, these assumptions are 
roughly equivalent to demanding that 
the source lies on the red line
in Fig.~\ref{figureeq}
if it is weakly absorbed (i.e., on the boundary of regions B and D), 
and on the yellow line if it is strongly absorbed
(i.e., on the boundary of regions C and D). This leads to a maximum brightness
temperature of a few times $10^{10}\,$K, as found by \cite{readhead94}.
Furthermore, as noted by \citet{readhead94}, such sources lie far from the 
threshold temperature, achieved along the dotted line in Fig.~\ref{figureeq}.

Replacing the assumption that the source flux is measured at the
synchrotron peak, by the requirement that its spectrum be flat, i.e., that it
lie in region~B of Fig.~\ref{figureeq}, one sees that a wide range of
brightness temperatures is available for sources in equipartition, extending
up to the threshold temperature found by \cite{kellermannpaulinytoth69}. 
Thus, the observed temperature distribution is not explained by the assumption
of equipartition. 

\section{Time dependence and acceleration}
\label{timedependent}

In order to explain the occurrence of brightness temperatures
above $10^{12}\,$K, \cite{slysh92} formulated a model involving a
monoenergetic electron distribution in a strongly absorbed source, in
the sense that $\gamma<\gamma_{\rm c}$, where $\gamma_{\rm c}$ is 
defined in Eqs.~(\ref{strongweak}) and (\ref{strongweak2}).  He
considered two scenarios, (i) a time-dependent one in which electrons
were injected at arbitrarily high Lorentz factors and allowed to cool
and (ii) one in which a strong continuous re-acceleration of the
electrons led to a high brightness temperature equilibrium.

In each case, the assumption that the source is strongly absorbed
leads to extreme values of the parameters. For example, in the first
scenario in which high energy particles are injected into the source,
\cite{slysh92} finds that a brightness temperature of $T_{\rm
B}>5\times10^{15}\,$K can be sustained over 1~day at an observing
frequency of $1\,$GHz. This is clearly in conflict with our
analysis. The electron Lorentz factor required to achieve this
temperature is $\gamma>10^5$. However, the condition that the source
is strongly absorbed, which is used in this model to estimate the
cooling rate, combined with the condition $\nusynch\approx1\,$GHz
required for a flat spectrum, leads to an extremely large Thomson
optical depth, $\tau\approx130$, as well as an implausibly low
magnetic field $B\approx 2\times10^{-11}\,$G. The parameter $\xi$ that
determines the inverse Compton luminosity is approximately $10^{12}$,
which implies an extremely large compactness of the inverse Compton
radiation from the source. The resulting copious pair production
invalidates the analysis and, ultimately, reduces the brightness
temperature achievable in the radio range. The same criticism applies
also to the second scenario described by \cite{slysh92} in which
acceleration balances inverse Compton losses to provide a brightness
temperature of $10^{14}\,$K at $1\,$GHz.

In the absence of Klein-Nishina effects on the scattering cross
section, we find the time dependence of the particle and photon
energies can be described by the three ordinary differential equations
(\ref{explosive}), (\ref{explosivea}) and
(\ref{electroneqa}). Inspection of these shows that if the threshold
temperature is exceeded ($\gamma>\gammaexp$), the inverse Compton
luminosity grows in a timescale of roughly the light-crossing time of
the source.  Thus, the threshold can only be substantially exceeded if
the acceleration process in Eq.~(\ref{electroneqa}) operates on a
shorter timescale.  However, these equations employ a spatial average
over the emission region.  Although a rapid acceleration rate might be
achieved locally in small regions of the source, once an average is
taken, no timescale in the system can be shorter than the light-crossing 
time of the region over which the accelerated particles are
distributed. In this case, the threshold temperature cannot be
significantly exceeded.

At first sight, Klein-Nishina effects offer a possible escape from this
conclusion. If even the first order scattering is suppressed, which
requires extremely large Lorentz factors for the electrons
($\gamma>10^{10}$ is needed for Klein-Nishina effects when scattering
$10\,$GHz photons), the strong reduction in the rate of cooling by
inverse Compton scattering suggests that higher brightness temperatures
$T_{\rm B}$ might be possible.

This is, however, not the case, because the rate of production of
electron-positron pairs by photon-photon interactions becomes
important.  The strength of this effect, which is not included in our
model equations, can be measured in terms of the \lq\lq compactness
parameter\rq\rq\ $\ell$ \citep[see, for
example][]{mastichiadiskirk95}, defined as
\eqb 
\ell&=&{\sigma_T RU_{N_{\rm max}}\over
h\nu_{N_{\rm max}}} 
\eqe 
where $U_{N_{\rm max}}$ is defined in Eq.~(\ref{timedepui}), 
$N_{\rm max}$ in Eq.~(\ref{defnmax}), and
$\nu_{N_{\rm max}}$ is taken to be $(4\gamma^2/3)^{N_{\rm
max}}\nu_0$. When $\ell>1$, one expects the pair-production rate to be
roughly equal to the light-crossing time of the source. This leads to
a sharp rise in the 
Thomson optical depth, invalidating the assumption of scatter-free
escape of synchrotron photons that is implicit in our model. The
associated confinement of these photons reduces the brightness
temperature.

\begin{figure}
\resizebox{\hsize}{!}{%
\includegraphics*[width=8cm]{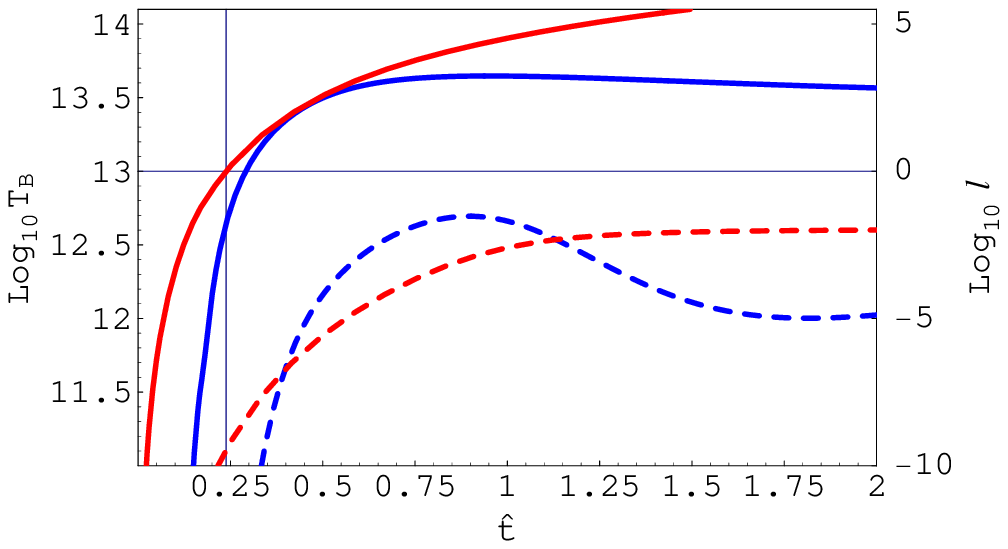}}
\resizebox{\hsize}{!}{%
\includegraphics*[width=8cm]{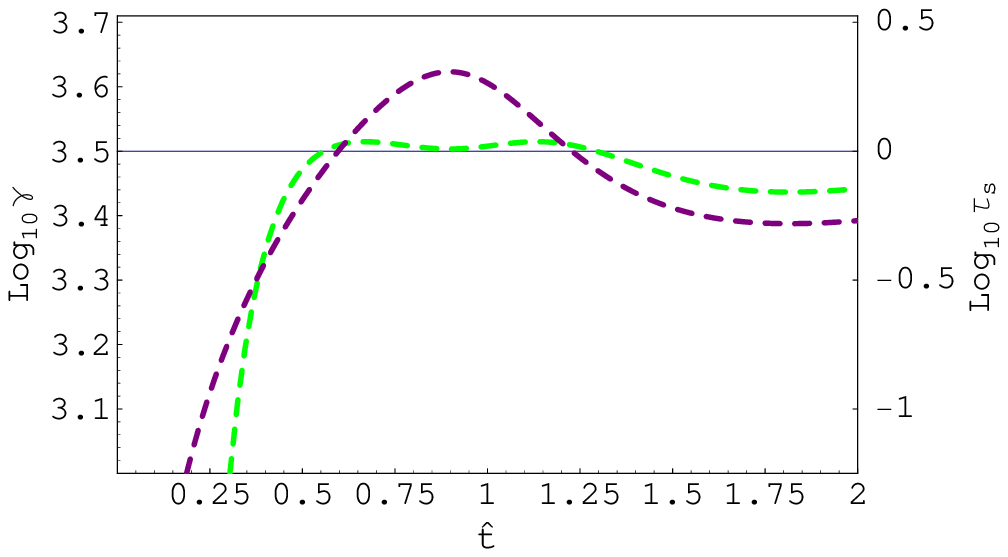}}
\caption{\label{figurestrg}{\em Upper panel\/}: 
The brightness temperature $\TB$ (blue), and the compactness $\ell$ (red) as functions of time, 
for two stationary, local sources ($\doppler=1$, $z=0$) with linear 
size $R=0.01$ pc, observed at $1\,$GHz. The Thomson optical depth
is $\tauT=0.01$ (dashed lines) and $\tauT=1$ (solid lines) and the 
remaining parameters are chosen such that the optical depth to synchrotron self-absorption $\taus\approx1$ at $\gamma=\gammac$ (see Eqs.~(\ref{strongweak}) and (\ref{strongweak2})). A horizontal line is drawn to indicate $\ell=1$. {\em Lower panel\/}: The electron Lorentz factor (purple dashed) and 
the optical depth to synchrotron self-absorption $\taus$ (green dashed)
for the case $\tauT=0.01$. A horizontal line indicates $\taus=1$.}
\end{figure}

We illustrate this in Fig.~\ref{figurestrg}, where we compare two
models with the same linear size $R$  (and observing frequency), 
but different electron densities $\nelec$ and different values of $B$, 
chosen as follows: For any given set of parameters, $R$, $B$ and $\nelec$, and observing
frequency $\nu_{\rm obs}$, the optical depth to synchrotron absorption
$\taus$, as defined in Eq.~(\ref{tau}), has a single maximum as
a function of $\gamma$, located close to the point where $\nu_{\rm
obs}$ equals the characteristic synchrotron frequency. 
If the source is optically thick to absorption at this
point, then $\gamma<\gamma_{\rm c}$, as described in
Sect.~\ref{synchrotron}, and the brightness temperature is roughly
$3\gamma m c^2/4k_{\rm B}$. If, on the other hand, the source is optically
thin at this point, then $\gamma>\gamma_{\rm c}$, but the brightness
temperature, given approximately 
by $\taus\times3\gamma m c^2/4k_{\rm B}$, decreases
to higher $\gamma$, as can be seen from Eq.~(\ref{tau}). Thus,
assuming inverse Compton scattering does not intervene, the maximum
brightness temperature is observed at a frequency such that $\tau_{\rm
s}\approx1$, when $\gamma=\gamma_{\rm c}$, which implies $x\approx1$. 
These conditions are imposed on the parameters of the 
models presented in Fig.~\ref{figurestrg}. In addition to the source 
size, chosen to be $R=0.01\,$pc and the observing frequency, set to 
$1\,$GHz, this leaves one 
free parameter, which we choose to be the optical depth to
Thomson scattering $\tauT$.

The upper panel in 
Fig.~\ref{figurestrg} shows the time-dependence of the 
brightness temperature found by solving
Eqs.~(\ref{timedepui}) and (\ref{electroneqb}) numerically for 
sources with $\tauT=0.01$ (dashed blue line) and 
$\tauT=1$ (solid blue line), without allowance
for Doppler boosting ($\doppler=1$). 
These sources have $\gammac=10^{3.6}$ and
$\gammac=10^{4.3}$, respectively and, in
the absence of inverse Compton cooling, they 
could potentially achieve brightness temperatures of 
$T_{\rm B}\approx10^{13.2}\,$K and
$T_{\rm B}\approx10^{13.9}\,$K. 
In order to do so, rapid acceleration is required, since 
for these source parameters, inverse Compton cooling 
leads to a time-asymptotic value of the Lorentz factor 
that is somewhat lower than $\gammac$ for slow acceleration. The exact value 
of the asymptotic solution depends on the strength of the acceleration. For acceleration 
on the light-crossing timescale, it corresponds to $a\approx\gamma/\gammatrap$
[see Eq.~(\ref{electroneqa})]. In Fig.~\ref{figurestrg} we choose 
  $a=1.5\gamma/\gammatrap$, which leads to an overshoot that slightly
exceeds $\gammac$. 

For $\tauT=0.01$, the compactness, shown  as a function of time
by the red dashed line, remains well below unity, so that the effects of pair
production can be neglected. However, this is not the case for
$\tauT=1$. Here, the compactness (solid red line) 
rises rapidly, reaching unity at
$\hat{t}\approx0.25$, where $T_{\rm B}\approx3.5\times10^{12}\,$K, 
well below its potential maximum. Thus, the attempt to gain higher brightness 
temperature by increasing $\tauT$, and, hence, $\gammac$, leads to a breakdown
in the model assumptions due to pair production. 

The lower panel of Fig.~\ref{figurestrg} shows the electron Lorentz factor and
the optical depth to synchrotron self-absorption
$\taus$ as functions of time for the case $\tauT=0.01$. The Lorentz
factor overshoots both its time-asymptotic value and $\gammac$.
Correspondingly, the optical depth, (shown as the green
dashed line) which initially rises with $\gamma$, 
reaching unity at $\gamma=\gammac$ goes through a maximum very shortly
afterwards. However, the overshoot is not sufficient to push $\taus$ back
below unity, and the maximum brightness temperature, which coincides with the
maximum Lorentz factor, remains at $T_{\rm B}=5\times10^{12}\,$K, 
somewhat below the value of $T_{\rm B}\approx10^{13.2}\,$K,
estimated for large optical depth.

\section{Conclusions}
\label{conclusions}

The well-known upper limit on the brightness temperature 
of a synchrotron source $T_{\rm B}\lesssim 10^{12}\,$K
imposed by the inverse Compton catastrophe, has been
reassessed, assuming a monoenergetic electron distribution. 

In weakly absorbed sources (see Eq.~(\ref{strongweak})), 
this distribution 
mimics the situation in which the conventional power-law 
is truncated to lower energies at a Lorentz factor $\gamma_{\rm min}$.
Using the standard theory of synchrotron emission and self-absorption, 
we find that, for such sources, the brightness temperature at a frequency of a
few GHz can reach approximately $10^{14}\,$K, the precise limit being given in
Eq.~(\ref{bright1}). Physically, this increased limit reflects the absence of
cool electrons in monoenergetic distributions and 
in those that are 
truncated or hard below a certain Lorentz factor. 
As a consequence, intra-day variable
sources can in principle be understood without recourse to other
mechanisms such as 
unusually large Doppler factors \citep{rees67}, 
coherent emission \citep[e.g.,][]{begelmanetal05} or
proton synchrotron radiation \citep{kardashev00}.

The possibility of exceeding the new limit in a time-dependent solution by
balancing losses against a strong acceleration term has been investigated
using a set of spatially averaged equations. Provided the acceleration process
remains causal i.e., the acceleration time averaged over the source remains
longer than the light-crossing time, we find a modest overshoot is
possible, but the maximum temperature is still restricted
by Eq.~(\ref{bright1}). In strongly absorbed sources, such as those
considered by \cite{slysh92}, high brightness temperatures cannot be attained
in a self-consistent model of the kind we discuss. The underlying reason is
that extremely compact sources would be required, in which copious
pair-production must be taken into account.
 
We have examined in detail the parameter space available to homogeneous
synchrotron sources of fixed size.
In the case of flat spectrum sources, we find that the imposition of the
condition of equipartition between the particle and magnetic field energy
densities does not result in a lower limit on the brightness temperature than
that given by the inverse Compton catastrophe. Suggestions to 
the contrary \citep{readhead94} are based on the more restrictive twin
assumptions that the power-law electron distribution is not 
truncated within the relevant range, and that the temperature is measured at
the point where the optical depth of the source is approximately unity. 
Consequently, the observed temperature distribution does not support the
equipartition hypothesis. We also find that flat spectrum sources
close to equipartition 
can approach the threshold temperature of the inverse Compton catastrophe, in
contrast with the finding based on the more restrictive assumptions in
\citet{readhead94}.

A corollary of the theory presented here is that for very bright sources,
which are necessarily weakly absorbed, 
a high degree of intrinsic circular polarisation is predicted
\citep{kirktsang06}. In addition, the theory can be tested by 
comparison of the predicted synchrotron 
spectrum with simultaneous observations of 
high brightness temperature sources in the radio to infra-red range 
and comparison of the predicted inverse Compton emission 
with measurements at MeV to GeV energies.

\appendix
\section{Synchrotron formulae}
\label{synchformulae}
We consider a region of homogeneous magnetic field $B$, linear dimension $R$,
(and volume $R^3$) containing monoenergetic electrons/positron of number
density $\nelec$ and Lorentz factor $\gamma$. The volume emissivity for
synchrotron radiation, summed over polarisations, is:
\eqb
j_\nu&=& 
{\sqrt{3}\over 4\pi}\nelec\alpha_{\rm f}\,
\hbar\Omega_{\rm L}\,\sin\theta\, F\left(x\right)
\label{emissivity}
\\  
x&=&\nu/\nu_{\rm c}
\label{xdef}
\\
\nu_{\rm c}(\gamma,\theta)&=&{3\Omega_{\rm L}\sin\theta\gamma^2\over4\pi}
\nonumber\\
&=&\nu_0\gamma^2\qquad[\nu_0=3\Omega_{\rm
    L}\sin\theta/(4\pi)]
\label{nusynchdef}
\\
F(x)&=&x\int_x^\infty \diff t K_{5/3}(t)
\label{fxdef}
\eqe
where $\alpha_{\rm f}=e^2/\hbar c$ is the fine-structure constant,
$\Omega_{\rm L}=eB/mc$ the Larmor frequency and $\theta$ the angle between the
magnetic field and the direction of the emitted radiation.
For small and large $x$ the limiting forms are
\eqb
F(x)&\approx&
{4\pi\over\sqrt{3}\Gamma(1/3)}\left({x\over2}\right)^{1/3}
\quad\textrm{for\ }x\ll1
\label{fsmallx}
\\
F(x)&\rightarrow&
\sqrt{\pi x\over2}\textrm{e}^{-x}
\quad\textrm{for\ }x\rightarrow\infty
\eqe

The absorption coefficient for unpolarised radiation is
\eqb
\alpha_\nu&=&{1\over2\sqrt{3}}{\nelec \sigmaT\over \alpha_{\rm f} 
b\sin\theta}{ K_{5/3}(x)\over\gamma^5}
\label{abscoeff}
\eqe
where $b=\hbar\Omega_{\rm L}/mc^2$ is the magnetic field in units of the
critical field $B_{\rm c}=4.414\times10^{13}\,$G and $\sigmaT$ is the Thomson
  cross-section. The limiting forms are:
\eqb
K_{5/3}(x)&\approx&
{2^{2/3}\Gamma(5/3)\over x^{5/3}}
\quad\textrm{for\ }x\ll1
\label{ksmallx}
\\
K_{5/3}(x)&\rightarrow&
\sqrt{\pi\over 2x}\textrm{e}^{-x}
\quad\textrm{for\ }x\rightarrow\infty
\eqe

Because $\alpha_\nu$ is a monotonically decreasing function of $x$, we can
define a unique $\xabs(b,\gamma)$ where the optical depth
$\taus=R\alpha_\nu$
for synchrotron absorption along a path of length $R$ is unity:
\eqb
R\alpha_\nu\left(\xabs\right)\,=\,1
\eqe
If $\xabs\ll1$, we have {\em weak absorption}
and for  $\xabs\gg1$ {\em strong absorption}.
The transition between the two regimes occurs near Lorentz factor 
$\gammac$, defined as
\eqb
\gammac&=&\left({\tauT\over 2\sqrt{3}\alpha_{\rm f}b\sin\theta}\right)^{1/5} 
\label{gammacdef}
\eqe
so that 
\eqb
\taus&=&\ghat^{-5}K_{5/3}(x)
\nonumber \\
&=&\frac{\sqrt{3}\tauT mc^3K_{5/3}(x)}{8\pi e^2\nu_{\rm c}\gamma^3}
\label{tau}
\eqe
where $\ghat=\gamma/\gammac$.
In the case of weak absorption, 
\eqb
\xabs&\approx& 2^{2/5}\left[\Gamma(5/3)\right]^{3/5}/\ghat^3
\quad\textrm{for\ }\ghat\gg1
\eqe
whereas in the strong absorption regime
\eqb
\xabs&\sim&-5\ln\ghat
\quad\textrm{for\ }\ghat\ll1
\eqe

Taking account of synchrotron emission and absorption and ignoring the role of
polarisation, the specific intensity in a direction which cuts the source 
on a path of length $R$ is
\eqb
I_\nu&=&S_\nu\left[1-\textrm{exp}\left(-\taus\right)\right]
\label{specintappendix}
\eqe
where the {\em source function} $S_\nu=j_\nu/\alpha_\nu$ is
\eqb
S_\nu&=&
\left({B^2\over 8\pi}\right)\left({9e^2\gammac^5\over2\pi mc^2}\right)
\sin^2\theta\, S(\ghat,x)
\label{sourcefunction}
\eqe
with 
\eqb
S(\ghat,x)&=&
{\ghat^5F(x)\over K_{5/3}(x)}
\\
&\rightarrow&
\left\lbrace
\begin{array}{ll}
{2\pi\over\sqrt{3}\Gamma(1/3)\Gamma(5/3)}\ghat^5 x^2 &\textrm{as\ }
x\rightarrow0 \\
\ghat^5 x &\textrm{as\ }x\rightarrow\infty
\end{array}
\right.
\eqe
and the optical depth to synchrotron absorption $\taus$ 
is a function of $\ghat$ and $x$.

To find the energy density $U_{\rm s}$
in synchrotron photons in a given source, 
$I_\nu$ must be integrated over angles and 
over frequency. The result depends on the geometry and 
optical depth as well as the position
within the source. However, an average value can be estimated by 
introducing a geometry dependent factor $\zeta\approx 1$:
\eqb
U_{\rm s}&\approx&{4\pi \zeta\over c}\int_0^\infty \diff\nu 
\left<I_\nu\right>
\label{xidef2}
\eqe
and denoting by $\left<I_\nu\right>$ the specific intensity evaluated at
$\theta=\pi/2$. Then 
\eqb
U_{\rm s}
&=&
\zeta\left({B^2\over 8\pi}\right)\
\left({27\alpha_{\rm f}\over 2\pi}\right)
b\gammac^7 U(\ghat)
\eqe
with
\eqb
U(\ghat)&=&
\ghat^2\int_0^\infty \diff x
S(\ghat,x)\left\lbrace 1-\textrm{exp}
\left[-\taus(\ghat,x)\right]\right\rbrace
\eqe
This integral 
is dominated by the region $x\gg \xabs$ 
in the weak absorption regime:
\eqb
U(\ghat)&\approx&\int_0^\infty \diff x\, S\taus
\nonumber\\
&=&\int_0^\infty \diff x\,F(x)
\nonumber\\
&=&{8\pi\ghat^2\over 9\sqrt{3}}\quad\textrm{for\ }\ghat\gg1
\label{weakabs}
\eqe
and by the region around $x=\xabs$ in the strong absorption regime:
\eqb
U(\ghat)&\approx&\int_0^{\xabs} \diff x\, S
\nonumber\\
&\approx&12.5\ghat^7\left(\ln\ghat\right)^2
\quad\textrm{for\ }\ghat\ll 1
\eqe
which suggests the simple approximation
\eqb
U(\ghat)&\approx&{12.5\ghat^7\over 
\left[0.183+\left(\ln\ghat\right)^{2}\right]^{-1}+7.75\ghat^{5}}
\label{simpleapprox}
\eqe
where the constant $0.183$ was chosen such that the approximation passes
through the point $U(1)=0.945$ found by numerical integration.
\bibliographystyle{aa}
\bibliography{6502asph}
\end{document}